\documentclass[conference]{IEEEtran}
\IEEEoverridecommandlockouts
\usepackage{cite}
\usepackage{pifont}
\usepackage{subcaption}
\usepackage{amsmath,amssymb,amsfonts}
\usepackage{algorithmic}
\usepackage{algorithm}
\usepackage{graphicx}
\usepackage{textcomp}
\usepackage{xcolor}
\usepackage{comment}
\usepackage{array}
\usepackage{booktabs}
\usepackage{makecell}
\usepackage{multirow}
\usepackage{mathtools}
\usepackage{threeparttable}

\IEEEoverridecommandlockouts%\IEEEpubid{\makebox[\columnwidth]{ 979-8-3503-0322-3/23/\$31.00 $\copyright$2023 IEEE \hfill}\hspace{\columnsep}\makebox[\columnwidth]{ }}

\begin{document}
\newtheorem{definition}{\it Definition}%[section]
\newtheorem{theorem}{\bf Theorem}%[section]
\newtheorem{lemma}{\it Lemma}
\newtheorem{corollary}{\it Corollary}
\newtheorem{remark}{\it Remark}
\newtheorem{example}{\it Example}
\newtheorem{case}{\bf Case Study}
\newtheorem{assumption}{\it Assumption}
\newtheorem{property}{\it Property}
\newtheorem{proposition}{\it Proposition}
% ===only use in IEEE format : END ====

\newcommand{\hP}[1]{{\boldsymbol h}_{{#1}{\bullet}}}
\newcommand{\hS}[1]{{\boldsymbol h}_{{\bullet}{#1}}}

\newcommand{\ba}{\boldsymbol{a}}
\newcommand{\baq}{\overline{q}}
\newcommand{\bA}{\boldsymbol{A}}
\newcommand{\bb}{\boldsymbol{b}}
\newcommand{\bB}{\boldsymbol{B}}
\newcommand{\bc}{\boldsymbol{c}}
\newcommand{\bcO}{\boldsymbol{\cal O}}
\newcommand{\bh}{\boldsymbol{h}}
\newcommand{\bH}{\boldsymbol{H}}
\newcommand{\bl}{\boldsymbol{l}}
\newcommand{\bm}{\boldsymbol{m}}
\newcommand{\bn}{\boldsymbol{n}}
\newcommand{\bo}{\boldsymbol{o}}
\newcommand{\bO}{\boldsymbol{O}}
\newcommand{\bp}{\boldsymbol{p}}
\newcommand{\bq}{\boldsymbol{q}}
\newcommand{\bR}{\boldsymbol{R}}
\newcommand{\bs}{\boldsymbol{s}}
\newcommand{\bS}{\boldsymbol{S}}
\newcommand{\bT}{\boldsymbol{T}}
\newcommand{\bu}{\boldsymbol{u}}
\newcommand{\bv}{\boldsymbol{v}}
\newcommand{\bw}{\boldsymbol{w}}

\newcommand{\balpha}{\boldsymbol{\alpha}}
\newcommand{\bbeta}{\boldsymbol{\beta}}
\newcommand{\bOmega}{\boldsymbol{\Omega}}
\newcommand{\bTheta}{\boldsymbol{\Theta}}
\newcommand{\bphi}{\boldsymbol{\phi}}
\newcommand{\btheta}{\boldsymbol{\theta}}
\newcommand{\bvarpi}{\boldsymbol{\varpi}}
\newcommand{\bpi}{\boldsymbol{\pi}}
\newcommand{\bpsi}{\boldsymbol{\psi}}
\newcommand{\bxi}{\boldsymbol{\xi}}
\newcommand{\bx}{\boldsymbol{x}}
\newcommand{\by}{\boldsymbol{y}}

\newcommand{\cA}{{\cal A}}
\newcommand{\bcA}{\boldsymbol{\cal A}}
\newcommand{\cB}{{\cal B}}
\newcommand{\cE}{{\cal E}}
\newcommand{\cG}{{\cal G}}
\newcommand{\cH}{{\cal H}}
\newcommand{\bcH}{\boldsymbol {\cal H}}
\newcommand{\cK}{{\cal K}}
\newcommand{\cO}{{\cal O}}
\newcommand{\cR}{{\cal R}}
\newcommand{\cS}{{\cal S}}
\newcommand{\dcS}{\ddot{{\cal S}}}
\newcommand{\ds}{\ddot{{s}}}
\newcommand{\cT}{{\cal T}}
\newcommand{\cU}{{\cal U}}
\newcommand{\wt}[1]{\widetilde{#1}}

\newcommand{\mA}{\mathbb{A}}
\newcommand{\mE}{\mathbb{E}}
\newcommand{\mG}{\mathbb{G}}
\newcommand{\mR}{\mathbb{R}}
\newcommand{\mS}{\mathbb{S}}
\newcommand{\mU}{\mathbb{U}}
\newcommand{\mV}{\mathbb{V}}
\newcommand{\mW}{\mathbb{W}}

\newcommand{\uq}{\underline{q}}
\newcommand{\ubq}{\underline{\boldsymbol q}}

\newcommand{\red}[1]{\textcolor[rgb]{1,0,0}{#1}}
\newcommand{\gre}[1]{\textcolor[rgb]{0,1,0}{#1}}
\newcommand{\blu}[1]{\textcolor[rgb]{0,0,1}{#1}}

\title{Physical-Layer Semantic-Aware Network for Zero-Shot Wireless Sensing}

% \author{
% Huixiang~Zhu, Yong~Xiao, \IEEEmembership{Senior~Member,~IEEE}, Yingyu Li, Guangming~Shi, \IEEEmembership{Fellow, IEEE},
% Walid Saad, \IEEEmembership{Fellow, IEEE}}

\author{\IEEEauthorblockA{Huixiang Zhu\IEEEauthorrefmark{1}, Yong~Xiao\IEEEauthorrefmark{1}\IEEEauthorrefmark{2}\IEEEauthorrefmark{3}, Yingyu Li\IEEEauthorrefmark{4}, Guangming~Shi\IEEEauthorrefmark{2}\IEEEauthorrefmark{5}\IEEEauthorrefmark{3}, Walid Saad\IEEEauthorrefmark{6} %, Yi Zhong\IEEEauthorrefmark{1}, Tao Han\IEEEauthorrefmark{1}\\
\IEEEauthorblockA{\IEEEauthorrefmark{1}School of Elect. Inform. \& Commun., Huazhong Univ. of Science \& Technology, China}
\IEEEauthorblockA{\IEEEauthorrefmark{2}Peng Cheng Laboratory, Shenzhen, China}
\IEEEauthorblockA{\IEEEauthorrefmark{3}Pazhou Laboratory (Huangpu), Guangzhou, China}
\IEEEauthorblockA{\IEEEauthorrefmark{4}School of Mech. Eng. and Elec. Info., China University of Geosciences(Wuhan), China}
\IEEEauthorblockA{\IEEEauthorrefmark{5}School of Artificial Intelligence, Xidian University, Xi'an, China}
\IEEEauthorblockA{\IEEEauthorrefmark{6}Bradley Department of Electrical and Computer Engineering, Virginia Tech, VA, USA}
}
\thanks{*This paper is accepted at IEEE International Conference on Network Protocols (ICNP) Workshop, Reykjavik, Iceland, October 10-13, 2023.}
}
%, Diep N. Nguyen, \IEEEmembership{Senior~Member,~IEEE}, and Dinh Thai Hoang, \IEEEmembership{Senior~Member,~IEEE} %Dusit~Niyato, \IEEEmembership{Fellow, IEEE}, Walid Saad, \IEEEmembership{Fellow, IEEE}, Mehdi Bennis, \IEEEmembership{Fellow, IEEE}

% \IEEEauthorblockA{\IEEEauthorrefmark{1}School of Elect. Inform. \& Commun., Huazhong Univ. of Science \& Technology, Wuhan, China}
% %\IEEEauthorblockA{\{zhuhuixiang, jdzsuper, chh\_eic\}@hust.edu.cn}
% }
\maketitle

\begin{abstract}
Device-free wireless sensing has recently attracted significant interest due to its potential to support a wide range of immersive human-machine interactive applications. However, data heterogeneity in wireless signals and data privacy regulation of distributed sensing have been considered as the major challenges that hinder the wide applications of wireless sensing in large area networking systems. Motivated by the observation that signals recorded by wireless receivers are closely related to a set of physical-layer semantic features, in this paper we propose a novel zero-shot wireless sensing solution that allows models constructed in one or a limited number of locations to be directly transferred to other locations without any labeled data. We develop a novel physical-layer semantic-aware network (pSAN) framework to characterize the correlation between physical-layer semantic features and the sensing data distributions across different receivers. We then propose a pSAN-based zero-shot learning solution in which each receiver can obtain a location-specific gesture recognition model by directly aggregating the already constructed models of other receivers. We theoretically prove that models obtained by our proposed solution can approach the optimal model without requiring any local model training. Experimental results once again verify that the accuracy of models derived by our proposed solution matches that of the models trained by the real labeled data based on supervised learning approach.
\end{abstract}

\vspace{-0.2cm}

\section{Introduction}
RF signal-based wireless sensing has attracted significant interest recently due to its potential to enabling device-free movement detection and tracking in a wide range of applications including smart healthcare, urban sensing, and unmanned surveillance system. It has also been recognized as the key driver for emerging applications that require immersive contact-free human-machine interactions such as augmented reality/virtual reality (AR/VR) and Tactile Internet\cite{XY2018TactileInternet}. Recent results have already shown that, by detecting changes of RF signal propagation and reflection patterns caused by human body, it is possible to recognize a wide range of human actions and gestures such as falling, walking, sitting, etc. If sensing results collected from multiple receivers can be jointly utilized, more fine-grained human gestures such as human skeletons, hand gestures, and finger movement can be detected \cite{liu2019wireless}.

\begin{figure}[!ht]
     \centering
     % \begin{subfigure}[b]{0.22\textwidth}
     %     \centering         \includegraphics[width=1.1\textwidth]{Simulations/Motivation/Motivation_CSI_Patterns.eps}
     %     \caption{\bf }
     %     \label{CSI_Patterns}
     % \end{subfigure}
     % \hfill
     \begin{subfigure}[b]{0.22\textwidth}
         \centering
        \includegraphics[width=1\textwidth]{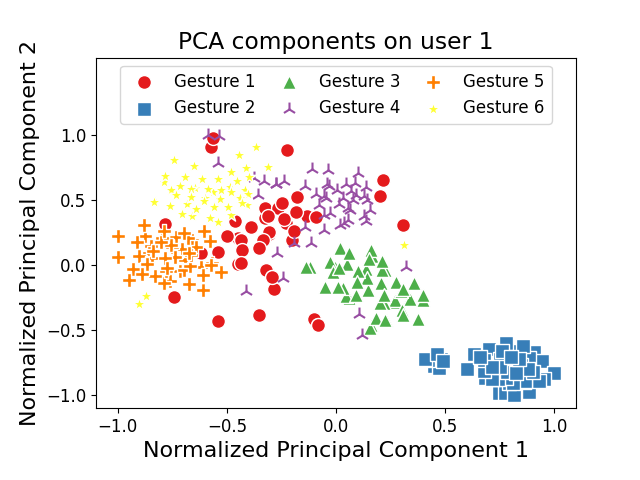}
         \caption{\bf }
         \label{dif_locations}
     \end{subfigure}
     \hfill
     \begin{subfigure}[b]{0.22\textwidth}
     \centering
     \includegraphics[width=1\textwidth]{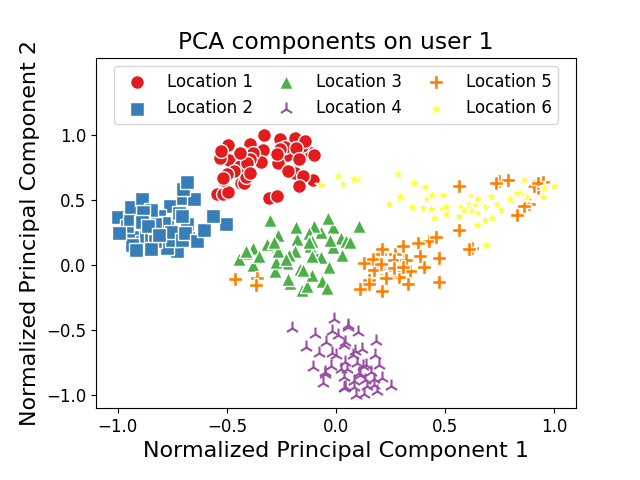}
         \caption{\bf }
         \label{dif_location}
     \end{subfigure}
    \caption{
    %(a) The CSI patterns of three different gestures collected at the same location. (b)
    (a) t-SNE-based visualization of statistical features of different gestures collected at the same location, and % between sensing data recorded at different locations.
    (b) visualization of statistical diversity of the same gestures collected at different locations.}
    \label{motivation_cluster}
    \vspace{-0.6cm}
\end{figure}

Most existing works adopt a
%focus on developing a centralized
one-fits-all modeling approach, in which a centralized model pre-trained based on the sensing data recorded from one or a limited number of locations, is deployed in a much wider range of locations and environments %any locations
throughout the considered area. % across a much wider range of locations and environments.
Unfortunately, it is known that the wireless signals, as well as their statistic features, exhibit highly temporal and spatial heterogeneity. More specifically, wireless signals are known to be highly specific to locations, environment, and human-related factors. For example, different locations of the transmitters, receivers, and users as well as room layouts will result in highly different wireless signal features and data distributions. Also, different body movement patterns as well as the orientations of the human users will also result in different spatial and temporal resolutions for gesture recognition. To shed more light on this observation, in Fig. \ref{motivation_cluster}, we compare the distributions of different sensing data recorded when the same human user performing different gestures at the same location (See Fig. \ref{motivation_cluster}(a)) as well as that when the same human user performing the same gesture at different locations (See Fig. \ref{motivation_cluster}(b)). We can observe that statistic features of sensing data can vary significantly at different locations. In other words, training a global model by combining sensing data recorded at different locations and ignoring the unique features of each individual environment and user profile will significantly reduce the wireless sensing accuracy and also result in high variations in sensing performance across different locations and gestures.

Another challenge for wireless sensing is it is quite difficult to collect high-quality labeled data samples. More specifically, as mentioned earlier, sensing signals recorded from different users at different locations generally have different distributions. In other words, the labeled data recorded at one location may not possess the unique statistical features in other locations. Also, due to the limit of physical space and cost, it is generally impossible to have a highly dense deployment of receivers and collect a large number of high-quality labeled samples to capture the spatial and temporal resolutions of different users and gestures in each considered area.      %different the deployment of the wireless receivers is generally limited by space and

To overcome the above challenges, we propose a zero-shot wireless sensing solution that can transfer the gesture recognition models trained in one or a limited number of locations to other locations with receivers that do not have any labeled data. Our proposed model is motivated by the observation that signals recorded by a wireless receiver is closely related to a set of physical-layer semantic features such as spatial layout, environment features, and human users' profiles. % channel between the transmitters and receivers, reflected and/or interrupted by
%human users physical environment and human users' body movements.
More specifically, we develop a novel physical-layer semantic-aware network (pSAN) framework to characterize the correlation between physical-layer semantic features and wireless signal distributions at different receivers. % and users' gestures and statistic features of the wireless sensing signals.
We then propose a pSAN-based zero-shot learning solution in which each receiver can obtain a location-specific gesture recognition model by directly aggregating the already trained models of other receivers. In our solution, the model aggregation coefficients are calculated based on the correlations between semantic features of different locations. We theoretically prove that the aggregated model obtained by our proposed solution can approach the optimal model without requiring any locally labeled data and local model training. Experiments conducted based on real-world wireless sensing datasets have been presented to verify our theoretical results.

We briefly summarize the key contributions as follows:
\begin{itemize}
\item We identify the physical-layer semantic features that determine the wireless signal distributions under different physical environments and users' gesture profiles. A novel pSAN framework is then proposed to capture the correlation between semantic features and the signal distributions at different receivers.
\item We develop a zero-shot learning solution based on pSAN which allows each receiver to obtain a location-specific model by linearly aggregating the  models trained by receivers. The semantic correlations between different locations have been utilized to calculate the model aggregation coefficient. Our proposed solution is simple to calculate and can obtain location-specific models across different areas with significantly reduced training dataset sizes and model training efforts.

\item We theoretically prove that the model obtained by our proposed solution can approach the optimal model in each specific location even without local model training.

\item Extensive experiments are conducted based on real-world wireless sensing datasets. Our results show that the models generated by our proposed model aggregation solutions can match models trained by real labeled data based on the supervised learning approach.
\end{itemize}

\section{Related Work}
\noindent
{\bf Wireless Sensing:}
In recent years, multiple wireless sensing techniques, including Wi-Fi, RFID, and Bluetooth, have been widely used to support various activity-aware sensing tasks.
Based on the different physical layer properties over wireless channels, commonly used wireless signal inputs can be broadly summarized into four categories, including received signal strength indicator (RSSI), channel state information (CSI), and Doppler frequency shift (DFS) \cite{liu2019wireless}. Although much effort has been invested in the in-depth study and practical application of these signal inputs with different physical characteristics, they all face significant challenges of spatiotemporal heterogeneity in complex wireless sensing environments \cite{li2021deep}.

\noindent
{\bf Semantic-Aware Network:} %It has already been observed
%Recent studies show that semantic features and correlations between different subjects and concepts can be helpful in inferring and interpreting implicit rules and principles\cite{X}.
The concept of semantic-aware network (SAN) was first proposed in \cite{XY2021SemanticCommMagazine} where it was defined as a novel architecture based on federated edge intelligence for supporting resource-efficient semantic-aware networking. The authors in \cite{xiao2022imitation} introduced a new multi-layer representation of semantic information that considers the hierarchical structure of implicit semantics. A graph-inspired structure was proposed in \cite{xiao2023reasoning} to represent the semantics of a message and convert its' graphical representation into a low-dimensional semantic space for efficient transmission.

\noindent
{\bf Zero-Shot Learning:} As an emerging learning paradigm, zero-shot learning has shown promising results %over traditional supervised learning approaches
%in applications that require to
in learning patterns and classifying instances in which the training and testing data are associated with different classes\cite{wang2019ZSLsurvey}.
%have not been seen previously
Existing zero-shot learning methods can be roughly divided into two categories: classification-based and instance-based methods. The former methods focus on learning instance classification models for new unseen scenarios/classes based on the known models\cite{palatucci2009zero}. % that identify unseen classes. % by leveraging knowledge from known classes.
The latter methods first try to establish labels for new unseen instances and then train supervised models based on the labeled instances\cite{lampert2009learning}. In this paper, we adopt the classification-based zero-shot learning methods for model transferring in which the wireless sensing models trained in some known environments can be transferred into new models that fit into new unseen physical environments. % based on the semantic correlations between environments.

%a model transferring solution is constructred based on the semantic correlations  maps .  % to  these instances belonging to the unseen classes and use them for classifier learning. %The key difference between them is the way the classifiers are learned.

\section{Network Model and Problem Formulation}

\subsection{Network Model}

We consider human gesture recognition based on a wireless sensing system consisting of multiple Wi-Fi transmitters and receivers deployed across different locations in the considered area.
Each receiver records wireless signals that are reflected and scattered by human users when performing a set of gestures. We focus on the decentralized sensing scenario in which each receiver stores its recorded wireless signals locally and, due to the constraints in data privacy, these signals cannot be sent to others. %Since the labeled datasets are expensive and limited. We therefore
We assume that only a limited number of receivers can have labeled wireless sensing data. Each of these receivers can then construct a location-specific model to recognize different gestures of the human users based on its local sensing data. There are some other receivers that cannot obtain any labeled sensing data. For these receivers, they cannot construct any local models using traditional supervised learning approach. Also, due to the spatial heterogeneity of the wireless signals, these receivers also cannot directly copy other models trained by receivers with labeled data. % the gesture recognition model constructed by one receiver cannot be directly reused by other receivers at different locations.

\subsection{Physical-layer Semantic Features}
We however observe that the statistics of the wireless sensing results are closely related to the semantic features of the physical environment and the human users' profiles. Motivated by this observation, we investigate whether it is possible to develop a zero-shot learning solution that allows models trained by one of a limited number of receivers with labeled data to be  transferred to other receivers based on the semantic correlations between different receivers.

Let us first define the semantic features in wireless sensing networks %as the key factors
that may influence the distribution of the sensing data recorded at each receiver. % is closely related to the physical environment and specifications of the gesture classes.
%More specifically,
It is known that the wireless signal recorded by a receiver is mainly characterized by the CSI of the wireless links connecting the transmitter and receiver, reflected and scattered by the gesture-performing human users as well as the physical objects located along side of the channel. More specifically,
%due to the multipath effects in physical environments,
the CSI recorded at arrival time $t$, subcarrier frequency $\theta$, and antenna $a$ can be written as follows \cite{zheng2019Widar}:
\begin{eqnarray}\label{eq_ReceivedSignal}
 H\left(t, \theta, a \right)= \sum_{n\in {\cal L}_S} A_n e^{ \theta_{\pi} \tau_n(\theta, a)} +\sum_{m\in {\cal L}_M} A_m(t) e^{\theta_{\pi} \tau_m(t, \theta, a)},
\end{eqnarray}
where $\theta_{\pi} \coloneqq -j 2 \pi \theta$, ${\cal L}_S$ and ${\cal L}_M$ are sets of static path components and dynamic path components, respectively. For each propagation path $l$ ($l\in {\cal L}_S \cup {\cal L}_M$), $A_l$ and $\tau_l$ respectively denote its channel attenuation factor and propagation delay. Here dynamic path components refer to those signals reflected from the moving target while the static path component refers to the direct path signal and the reflection signal from static objects such as walls. Importantly, since CSI measurements are discrete in time (packet), frequency (subcarrier), and space (antenna) \cite{halperin2010predictable}, the shift $ \tau_l$ of the $l$th signal path in (\ref{eq_ReceivedSignal}) can be represented as follows:
\begin{eqnarray}
\tau_l\left(\theta, a \right)&=& \tau_0 + \Delta a_l \cdot \psi_0, \mbox{for } l \in {\cal L}_S \label{eq_signalphase_1} \\
 \tau_l\left(t, \theta, a \right)&=& \tau_0 -\frac{\rho_0}{\Delta \theta_l}\Delta t_l + \Delta a_l \cdot \psi_0 , \mbox{for } l \in {\cal L}_M  \label{eq_signalphase_2}
\end{eqnarray}
where $\Delta t_l$, $\Delta \theta_l$, $\Delta a_l$ are differences of packet, subcarrier and spatial position between $H\left(t, \theta, s \right)$. We define the reference path $H\left(0,0,0 \right)$ with the propagation delay $\tau_0$, DFS $\rho_0$ and AoA $\psi_0$.

From (\ref{eq_ReceivedSignal}), we can observe that the wireless sensing signals recorded by receiver $k \in  {\cal K}$ are closely related to the following two types of physical-layer semantic features:%, denoted , consisting of

\noindent
{\bf Physical environment-related semantics (P-semantics)}: include the semantic features related to the physical environment such as environmental layout and the relative locations and orientations among transmitter, receiver, and human user. According to (\ref{eq_ReceivedSignal}), we can write the feature vectors of P-semantics of a receiver $k$ as $\bu_k = \langle A_n, \tau_n, \psi_n \rangle_{n\in {\cal L}_S}$. %Let $\bar \bu_k$ be the embedding vector of $s_u(\bu_k) \in \mathcal{S} \equiv \mathbb{R}^b$.

\noindent
{\bf Gesture-related semantics (G-semantics)}: include the semantic features associated with gestures such as the users' body coordinate and movement patterns of gestures. %, relative orientation to the transmitter and receiver as well as the body coordinate and movement pattern of the gesture.
Similarly, according to (\ref{eq_ReceivedSignal}), we can write the feature vector of G-semantics of receiver $k$ as $\bv_k = \langle A_m, \tau_m, \psi_m, \rho_m \rangle_{m\in {\cal L}_M}$. %Let $\bar \bv_k$ be the embedding vector of $s_v(\bv_k) \in \mathcal{S} \equiv \mathbb{R}^b$.

We can combine both P- and G-semantics and write the physical-layer semantic features of wireless signals recorded by receiver $k$ as $\bphi_k = \langle \bu_k, \bv_k \rangle$. We can observe that the physical-layer semantic features are location and user specific and therefore each receiver $k$ has a unique semantic feature $\bphi_k$ which plays a key role in determining the probability distribution of the local wireless signal. %, as will be verified later.

\subsection{Physical-Layer Semantic-Aware Network}
Let us now formally introduce the concept of physical-layer semantic-aware network (pSAN) as follows:
\begin{definition}
A {\em physical-layer semantic-aware network} (pSAN) is a wireless sensing network in which the physical-layer semantic features, including both P-semantics and G-semantics, can be aware, or known, by each receiver.
\end{definition}

In pSAN, %the semantic features can be utilized by some receivers to establish local models based on the existing models trained by other receivers.
the correlations between semantic features of different receivers can be used to infer the correlations between different location-specific models associated with these receivers. % utilized by some receivers to estimate
More formally, let ${\cal K}'$ be the subset of $K'$ receivers that have labeled wireless sensing signals associated with a set of gesture classes. To simplify our description, we use $k'$ to denote the receiver with labeled local sensing data, i.e., $k'\in {\cal K’}$.  Let ${\cal D}_{k'}$ be the set of labeled sensing data at receiver $k'$. We assume the labeled data samples at different receivers in ${\cal K}’$  are associated with the same set of gesture classes. Similarly, we use ${\cal K}''$ to denote the set of $K''$  receivers without any labeled sensing data. Also, let $k''$ be the receiver that does not have any labeled data for $k'' \in {\cal K}''$. Let $\cal K$ be the set of all the $K$ receivers, i.e., ${\cal K} = {\cal K'} \cup {\cal K''}$.

%The key idea is to establish a low-dimensional semantic embedding space that can captures the correlations between the statistic features as well as the resulting gesture recognition models of sensing results at different receiver locations. More specifically, let $\bar\bphi_k$ be the $p$-dimensional semantic embedding of $\bphi_k$ for $k\in {\cal K}$. We also use $\Gamma \left( \bar\bphi_{k'}, \bar\bphi_{k''} \right)$ to denote y Two semantic embeddings $\bphi_{k'}$ and $\bphi_{k''}$ are close in the semantic embedding space if  % data models trained for recognizing the same set of gestures at different receiver locations.
The key idea is to establish a low-dimensional semantic embedding space that captures the correlations between the key statistic features that play a key role in determine the gesture recognition models based on the wireless signals recorded at different receiver locations. More specifically, let $\bar\bphi_k$ be the $p$-dimensional semantic embedding  of $\bphi_k$ for $k\in {\cal K}$. We also use $S \left( \bar\bphi_{k'}, \bar\bphi_{k''} \right)$ to denote the semantic distance between semantic embeddings $\bar \bphi_{k'}$ and $\bar \bphi_{k''}$.

Each labeled sensing data $\zeta_{k',i}(\bar\bphi_{k'}) = \langle x_{k',i} (\bar\bphi_{k'}), y_{k',i} \rangle$ recorded by receiver $k'$ consists of a signal feature vector $x_{k',i}(\bar\bphi_{k'})$, %, corresponding to a limited number of features
associated with each instance of CSI recorded by receiver $k'$ with semantic $\bar\bphi_{k'}$, and a class label $y_{k',i}$ that belongs to one of a set $\cal Y$ of gesture classes. Each receiver $k'\in {\cal K}'$ can then construct a local model by minimizing its local objective function,
\begin{eqnarray}
\lefteqn{\min_{\pmb{\omega}_{k'}(\bar\bphi_{k'})} F_{k'} \left(\pmb{\omega}_{k'}(\bar\bphi_{k'}) \right)} \nonumber \\
&& = {1 \over |{\cal D}_{k'}|} \sum_{\zeta_{k',i}(\bar\bphi_{k'}) \in {\cal D}_{k'}}\left[f_{k'}\left(\pmb{\omega}_{k'}(\bar\bphi_{k'}); \zeta_{k',i}(\bar\bphi_{k'}) \right)\right].
\end{eqnarray}

We then need to learn a mapping function that can infer the correlations between different receivers' models according to their semantic distance.

\subsection{Problem Formulation}
%As mentioned earlier, in wireless sensing, labeled data is expensive and important. Also, since
In our considered decentralized wireless sensing scenario, the data recorded by each receiver cannot be exposed to others. It is however possible for the receivers with labeled data to expose their local models to other receivers. In the rest of this paper, we will develop a pSAN-based model aggregation approach in which each receiver $k''\in {\cal K}''$ can construct a location-specific model by aggregating models trained by other receivers in ${\cal K}'$ based on the semantic correlations.

The main objective is to let the aggregated model at each receiver $k''$ be able to approach the local optimal model,
\begin{eqnarray}\label{main_objective}
\min_{\pmb{\omega}_{k''}} F_{k''} \left(\pmb{\omega}_{k''}(\bar\bphi_{k''}) \right) - F_{k''} \left(\pmb{\omega}^*_{k''} \right)
\end{eqnarray}
where
%More formally, each receiver $k''$ will establish local model
$\pmb{\omega}_{k''}$ is the aggregated model of receiver $k''$ calculated as follows:
\begin{eqnarray}\label{aggregate_euqa}
    \pmb{\omega}_{k''} = \sum_{k' \in {\cal K}'} \xi_{k', k''} \left( S ( \bar\bphi_{k'}, \bar\bphi_{k''} ) \right) \pmb{\omega}_{k'}
\end{eqnarray}
where $\xi_{k', k''} (\cdot)$ is a semantic-aware function that maps the semantic distance between receivers $k'$ and $k''$ to a normalized model aggregation coefficient value. We will give a more detailed discussion about $\xi_{k', k''} (\cdot)$ and prove the convergence result of our proposed  solution later.

% we aim at developing an optimization algorithm that can transfer models trained by receivers with labeled data to other receivers without any labeled data based on the relationship between physical-layer semantic features of different receivers.

%\newpage
\section{pSAN-based Zero-Shot Wireless Sensing}

%As mentioned earlier, the main objective is to
In this section, we will introduce a pSAN-based zero-shot learning algorithm that can transfer models trained by receivers with labeled data to other receivers based on the semantic correlations between different receivers.
%Our algorithm includes three key procedures: (1) collaborative local model training at receivers with labeled data; (2) learning a mapping function that maps the semantic distance between receivers to model correlations; %learning a non-linear function that maps the semantic distance to the similarity between personalized models; and (3) transferring the already trained models at some receivers to the target receivers based on semantic correlations. %In this section, we will present the details of these processes step by step.

\subsection{Model Training at Receivers with Labeled Data}
Different from traditional %federated learning-based
distributed  model training approaches that collaboratively train a globally shared model ${\boldsymbol \omega}$ to minimize the overall loss $\sum_{k'\in {\cal K}'} F_{k'}({\boldsymbol \omega})$, we decompose the overall goal of model construction into $K'$ individual local objective $ F_{k'}({\boldsymbol \omega_{k'}})$, each is locally trained at a receiver $k'$. This allows each individual receiver to train a personalized model $\pmb{\omega}_{k'}(\bar\bphi_{k'}) \in \mathbb{R}^d$ using its local labeled data.
To collaboratively train personalized models in parallel for all the receivers in ${\cal K}'$, we formulate the personalized model training problem as a  multi-task federated learning problem. Denote by $\mathcal{F}({\boldsymbol \Omega})=\sum_{k'\in \mathcal{K}'} F_{k'}({\boldsymbol \omega}_{k'})$ and $\mathcal{R}({\boldsymbol \Omega})=\sum_{{k'}\in \mathcal{K}'} \sum_{{k'} \neq \tilde{{k'}}} R\left(\| \pmb{\omega}_{k'}- \pmb{\omega}_{\tilde{{k'}}}\|^2\right)$ the first and the second terms of $\mathcal{J}({\boldsymbol \Omega})$, the problem can be written as
\begin{eqnarray}\label{global_obj}
& \min\limits_{{\boldsymbol \Omega}\in \mathbb{R}^{d\times {K'}}}&
\mathcal{J}({\boldsymbol \Omega}) \coloneqq
 \mathcal{F}({\boldsymbol \Omega})+\gamma \mathcal{R}({\boldsymbol \Omega}),
\end{eqnarray}
where $\pmb{\Omega}= [\pmb{\omega}_1, \cdots, \pmb{\omega}_{K'}]^{\top}$ to denote a $d$-by-${K'}$ dimensional collective model matrix that collects $\pmb{\omega}_1, \cdots, \pmb{\omega}_K$ as its columns, $\gamma >0$ is a regularization parameter, and $\|\cdot\|$ is the Euclidean norm. $R(\|\pmb{\omega}_{k'}- \pmb{\omega}_{\tilde{{k'}}}\|^2)$ %: \mathbb{R} \to \mathbb{R}$ denotes
is a non-linear attention-inducing function that maps model distances to a loss value. One commonly adopted function is the negative exponential function, given by  $1-e^{-\|\pmb{\omega}_{k'}- \pmb{\omega}_{\tilde{{k'}}}\|^2/{\sigma}}$ with hyperparameter $\sigma$.

To solve the optimization problem in (\ref{global_obj}), % $\mathcal{J}(\cdot)$,
we develop an incremental-type optimization scheme based on {\cite{bertsekas2011incremental}} that alternatively optimizes $\mathcal{F}(\cdot)$ and $\mathcal{R}(\cdot)$ until convergence. We assume the local model training at receivers in ${\cal K}'$ can be coordinated by a central server.
%Specifically, in
During the $t$th coordination round, each receiver $k' \in {\cal K}'$ first downloads the updated $d$-dimensional global model from the central server, i.e., $\pmb{\omega}^{t,0}_{k'}=\pmb{\psi}^{t}_{k'}$, where $\pmb{\omega}^{t,0}_{k'} \in \mathbb{R}^d$ is the initial personalized model of receiver $k'$ labeled as the $0$th local iteration in the $t$th coordination round. Then, each receiver $k$ solves the local objective by applying an unbiased gradient estimator (e.g., vanilla SGD) to perform $E$ local epochs of local training %on the received model $\pmb{\omega}^{t,0}_{k'}$
as follows:
\begin{equation}
\begin{aligned}
\pmb{\omega}^{t,e+1}_{k'} = \pmb{\omega}^{t,e}_{k'}-\eta \nabla \widetilde{F}_{k'}(\pmb{\omega}^{t,e}_{k'}), \; \mbox{for}\; e = 0, \ldots, E-1
\label{Local_SGD}
\end{aligned}
\end{equation}
where $\eta$ is the learning rate, $\nabla \widetilde{F}_{k'}(\pmb{\omega}^{t,e}_{k'})$ is the unbiased stochastic gradient of $\nabla F_{k'}(\pmb{\omega}^{t,e}_{k'}, \zeta^{t,e}_{k'}(\bar\bphi_{k'}))$ on the mini-batch $\zeta^{t,e}_{k'}(\bar\bphi_{k'})$ uniformly sampled from the local dataset $\mathcal{D}_{k'}$. After finishing $E$  epochs of local SGDs, each receiver will send the updated model $\pmb{\omega}^{t, E}_{k'}$ to the central server for model coordination. On the central server, a gradient descend step is applied to optimize  $\mathcal{R}(\cdot)$ as follows:
\begin{eqnarray}
 \pmb{\psi}^{t+1}_{k'} = \pmb{\omega}^{t,E}_{k'}- \sum_{{k'} \neq \tilde{k'} } \alpha \lambda\nabla R(\| \pmb{\omega}^{t,E}_{k'}- \pmb{\omega}^{t,E}_{\tilde{k'}}\|^2),
 \label{Server_gradient}
\end{eqnarray}
where $\alpha$ denotes the gradient step size. The above iterative processes repeat until a preset maximum number $T$ of coordination rounds  is reached.

\subsection{Mapping from Semantic Distance to Model Correlation}
The core idea of this paper is to directly construct an optimal model for each receiver $k'' \in {\cal K}''$ by aggregating models trained by other receivers in ${\cal K}'$ based on the semantic correlations. Therefore, it is significantly essential for us to establish the mapping between the semantic distance $S ( \bar\bphi_{k'}, \bar\bphi_{k''} )$ and the model similarity $M ( \pmb{\omega}_{k'}, \pmb{\omega}_{k''} )$ between receiver $k'$ and $k''$. %However, model similarity does not necessarily increase proportionally as the semantic distance decreases.
%For this reason, we deploy a deep feedforward network on the server as a non-linear mapping function $R(\cdot)$ to approximate the potential mappinadg relationship.

In order to learn a mapping function $G(\cdot)$, we first construct a set of training pairs $\{S ( \bar\bphi_{i'}, \bar\bphi_{j'} ), M ( \pmb{\omega}_{i'}, \pmb{\omega}_{j'} )\}$ ($i', j' \in {\cal K}'$), where $\bar\bphi_{i'}$ and $\pmb{\omega}_{i'}$ denote the $p$-dimensional semantic embedding of $\bphi_{i'}$ and the personalized model $\pmb{\omega}^{0,E}_{i'}$ uploaded from receiver $i$, respectively. Then, we employ a deep feedforward network to learn the mapping $y = G(x, \pmb{m})$, where $\pmb{m}$ is the parameter set of $G(\cdot)$. When using the calculus of variations to approximate the optimal $G(\cdot, \pmb{m}^{*})$, the target optimization problem can be presented as follows:
\begin{eqnarray}
\pmb{m}^{*} = \arg \min_{\pmb{m}} \sum_{\substack{j\neq i, \\ i,j \in {\cal K}'}}\|M ( \pmb{\omega}_{i'}, \pmb{\omega}_{j'} )-G(S ( \bar\bphi_{i'}, \bar\bphi_{j'} ), \pmb{m})\|^2,
\vspace{-0.5cm}
\end{eqnarray}
where we apply the same distance metric to calculate $S(\cdot)$ and $M(\cdot)$. For example, when applying cosine distance to measure both semantic distance and model distance, we have
$
S ( \bar\bphi_{i'}, \bar\bphi_{j'} )=\frac{{\bar\bphi_{i'}}^{\top}\bar\bphi_{j'}}{\|\bar\bphi_{i'}\|\cdot \|\bar\bphi_{j'}\|}, M (  \pmb{\omega}_{i'}, \pmb{\omega}_{j'} )=\frac{{\pmb{\omega}_{i'}}^{\top}\pmb{\omega}_{j'}}{\|\pmb{\omega}_{i'}\|\cdot \|\pmb{\omega}_{j'}\|}.
$
%After obtaining the optimal $J^*(\cdot)$, the server will perform semantic-aware model aggregation to customize the next-round model for each target receiver.

%However, model similarity does not necessarily increase proportionally as the semantic distance decreases. For this reason, we deploy a deep feedforward network on the server as a non-linear mapping function $R(\cdot)$ to approximate the potential mapping relationship.

\subsection{Model Transferring to Receivers without Labeled Data}

We can now construct a personalized model $\pmb{\omega}_{k''}$ for each receiver $k'' \in {\cal K}''$ by linearly aggregating the local models trained by receiver $k' \in {\cal K}'$, i.e., $\pmb{\omega}_{k''} = \sum_{k' \in {\cal K}'} \xi_{k', k''} \pmb{\omega}_{k'}$, where $\xi_{k', k''}$ is the normalized model aggregation coefficient. To improve the efficiency of model collaboration, higher values of model aggregation coefficients are assigned to models with smaller model distance as well as smaller semantic distance, i.e., the smaller the value of $M( \pmb{\omega}_{k'}, \pmb{\omega}_{k''})$, the higher the value of $\xi_{k', k''}$.

For receiver $k' \in {\cal K}'$, the gradient step (\ref{Server_gradient}) on the central server can be converted into a convex combination operation given by
\begin{equation}
\begin{aligned}
 \pmb{\omega}^{t+1}_{k'} = \sum_{\tilde{{k'}} \in {\cal K}'} \xi^t_{{k'},\tilde{{k'}}} \pmb{\omega}^{t,E}_{\tilde{{k'}}}
 %\xi^t_{{k'},1}\cdot \pmb{\omega}^{t,E}_1+ \cdots + \xi^t_{{k'},{K'}} \cdot \pmb{\omega}^{t,E}_{K'},
 \label{convex_combination}
\end{aligned}
\vspace{-0.2cm}
\end{equation}

where $\xi^t_{{k'},\tilde{{k'}}}$ is given by
\begin{equation}
\xi^t_{{k'},\tilde{{k'}}}=\left\{
\begin{aligned}
&2\alpha \lambda R^{'}(\|\pmb{\omega}^{t,E}_{k'}-
 \pmb{\omega}^{t,E}_{\tilde{{k'}}}\|^2) , & {k'} \neq \tilde{{k'}}, \\
&1- 2\alpha \lambda \sum^{K'}_{\tilde{{k'}} \neq {k'} } R^{'}(\|\pmb{\omega}^{t,E}_{k'}-
 \pmb{\omega}^{t,E}_{\tilde{{k'}}}\|^2), & {k'} = \tilde{{k'}}.
\end{aligned}
\right.
\label{333}
\end{equation}

In (\ref{convex_combination}), we have $\xi_{{k'},\tilde{{k'}}}\in \mathbb{R}_{+}$ for $ \tilde{k'} \in {\cal K}'$, $\sum^{K'}_{\tilde{{k'}}=1}\xi_{{k'},\tilde{{k'}}} = 1$, and $R^{'}(\cdot)$ denotes the derivation of $R(\cdot)$.
%$\xi^t_{{k'},\tilde{{k'}}}$ ($1 \le {k'} \le {K'}, 1 \le \tilde{{k'}} \le K'$) denotes the aggregate weight assigned to model $\pmb{\omega}^t_{\tilde{{k'}}}$ when aggregating the next-round intermediate matrix $\pmb{\omega}^{t+1}_{k'}$.

After establishing the mapping from the model distance to the normalized aggregation coefficient, the target model $\pmb{\omega}_{k''}$ for receiver $k'' \in {\cal K}''$ can be calculated as follows:
\begin{eqnarray}\label{aggregate_euqa_1}
    \pmb{\omega}_{k''} = \sum_{k' \in {\cal K}'} 2\alpha_{k'} R^{'}(G(S ( \bar\bphi_{k'}, \bar\bphi_{k''} ), \pmb{m}^{*})) \pmb{\omega}^{T-1}_{k'}.
\end{eqnarray}

\vspace{-0.4cm}
\begin{algorithm}[H]
	\caption{pSAN-based Zero-Shot Algorithm}\label{Algorithm}\scriptsize
	{\bf Input}: Target rounds $T$; Local SGD steps $E$; Number of receivers with labeled data $K'$; Number of receivers without labeled data $K''$; Datasets $\mathcal{D}_1$, $\dots$, $\mathcal{D}_{K'}$; Semantic features $\bphi_{1}$, $\dots$, $\bphi_{K}$. \\
	{\bf Output}: $\pmb{\omega}^{T-1}_0$, $\dots$, $\pmb{\omega}^{T-1}_{K'-1}$, and $\pmb{\omega}_0$, $\dots$, $\pmb{\omega}_{K''-1}$.
    \begin{itemize}
    \item[ 1:] Server broadcasts initial model $\pmb{\psi}_0$ to all sites;
    \item[ 2:] {\bf for} $t$ = 0, $\cdots$, $T-1$ {\bf do}
    \item[ 3:] \quad {\bf for} receiver $k' \in {\cal K}'$ {\bf in parallel do}
    \item[ 4:] \quad \quad {\bf for} $e$ = 0, $\cdots$, $E-1$ {\bf do}
    \item[ 5:] \quad \quad \quad Uniformly sample a mini-batch $\zeta^{t,e}_{k'}$ from $\mathcal{D}_{k'}$;
    \item[ 6:] \quad \quad \quad Perform SGD on $\pmb{\omega}^{t,e}_{k'}$ using (\ref{Local_SGD});
    \item[ 7:] \quad \quad {\bf end for}
    \item[ 8:] \quad \quad Upload $\pmb{\omega}^{t,E-1}_{k'}$ and $\bar\bphi_{k'}$ to the server;
    \item[ 9:] \quad {\bf end parallel for}
    \item[ 10:] \quad {\bf for} $k'$ = 0, $\cdots$, $K'-1$ {\bf do on server}
    \item[ 11:] \quad \quad Calculate next-round model $\pmb{\omega}^{t+1}_{k'}$ using (\ref{Server_gradient});
    \item[ 12:] \quad \quad Send $\pmb{\omega}^{t+1}_{k'}$ to receiver ${k'}$.
    \item[ 13:] \quad {\bf end for on server}
    \item[ 14:] {\bf end for}
    \item[ 15:] {\bf for} $k''$ = 0, $\cdots$, $K''-1$ {\bf do on server}
    \item[ 16:] \quad Calculate the target model $\pmb{\omega}_{k''}$ using (\ref{aggregate_euqa});
    \item[ 17:] \quad Send $\pmb{\omega}_{k''}$ to receiver ${k''}$.
    \item[ 18:] {\bf end for on server}
    \end{itemize}
\end{algorithm}

\vspace{-0.6cm}
\section{Theoretical Results}
\subsection{Assumptions and Key Lemma}
%To begin with, we make the following assumptions on the loss function $F_k$ for $k \in {\cal K} $. We will then present a key lemma that is helpful in analyzing the convergence of our algorithm with strongly convex and smooth $F_k$.

\begin{assumption}
($\mu$-convexity and L-smoothness) $F_1, \cdots, F_K$ are all $\mu$-convex and $L$-smooth: i.e.,$\frac{\mu}{2}\|\pmb{\nu}-\pmb{\omega}\|^2\le F_k(\pmb{\nu})-F_k(\pmb{\omega})-\langle \nabla F_k(\pmb{\omega}), \pmb{\nu}-\pmb{\omega}\rangle \le \frac{L}{2}\|\pmb{\nu}-\pmb{\omega}\|^2 $, for all $\pmb{\nu}, \pmb{\omega} \in \mathbb{R}^d$ and $k \in {\cal K}$.
\label{assumption1}
\end{assumption}

\begin{assumption}
(Bounded Variance) The variance of stochastic gradients in each receiver is bounded:   $\mathbb{E}\|\nabla \widetilde{F}_k(\pmb{\omega}_k, \zeta_k)-\nabla F_k(\pmb{\omega}_k)\|^2 \le \sigma^2_F$, for all $k \in {\cal K} $.
\label{assumption3}
\end{assumption}

\begin{assumption}
(Bounded Diversity) The stochastic gradient of $R(\cdot)$ is bounded: $\mathbb{E}\| R^{'}(\cdot)\|\le \kappa_R$, for all $k \in {\cal K} $.
\label{assumption4}
\end{assumption}

Assumption 1 is commonly adopted for many existing SGD-based convergence analysis. Assumptions 2 and 3 have already been shown to be reasonable for %is widely adopted in most federated convergence analyses and  has been verified in various
practical applications \cite{t2020personalized}. %Assumption 3 holds because that there is $\kappa_R=\frac {1}{2\alpha \lambda}$ to ensure $0\le \xi_{k,\tilde{k}} \le 1$ for all $ k \in \mathcal{K}$ and $\tilde{k} \in \mathcal{K}$.

\begin{lemma}
Suppose that Assumption 1 holds and $\gamma\rho >2L$. Let ${\boldsymbol \Omega^{*}}=[\pmb{\omega}^{*}_1, \cdots, \pmb{\omega}^{*}_K]$ denote the optimal solution of (\ref{global_obj}). There exists $\sigma^2_{F,1}$, e.g., $\sigma^2_{F,1}=\|\mathcal{F}(0)\|^2 \frac{2\gamma\rho}{\gamma\rho-2L}$,
\begin{equation}
\begin{aligned}
 \|\nabla \mathcal{F}({\boldsymbol \Omega})\|^2 \le \sigma^2_{F,2}+\frac{4L^2}{\mu}[\mathcal{J}({\boldsymbol \Omega})-\mathcal{J}({\boldsymbol \Omega}^{*})].
 \label{lemma1_2}
\end{aligned}
\end{equation}
\label{lemma1}
The bound is tight in the sense that $\sigma^2_{F,1}=0$ for the i.i.d. cases, while $\sigma^2_{F,1}>0$ for non-i.i.d. cases.
\end{lemma}

\begin{theorem}
 Suppose Assumptions \ref{assumption1}-\ref{assumption4} hold and $\gamma, \mu, L, \kappa_R, \sigma^2_{F}$ are defined therein. If $\gamma>2L$ and $T\ge \frac{4}{\tilde{\eta}_1 \mu}$, there exists learning rate $\eta\le\frac{\tilde{\eta}_1}{E}$ such that,
\begin{equation}
\begin{aligned}
\mathbb{E}[\mathcal{J}(\widetilde{\pmb{\Omega}}^{T})-\mathcal{J}(\pmb{\Omega}^{*})] \le& \mathcal{O}( \Delta_0 \mu e^{\frac{-\tilde{\eta}_1 \mu T}{4} } +\frac{M_1}{\mu^2T^2EK}+\frac{M_2}{\mu TEK}) ,
\label{theorem1}
\end{aligned}
\end{equation}
where $\widetilde{\pmb{\Omega}}^{T}=\sum^{T-1}_{t=0}\alpha_t \pmb{\Omega}_t/\sum^{T-1}_{t=0}\alpha_t$ with $\alpha_t \coloneqq(1-\tilde{\eta}_1\mu/4)^{-(t+1)}$, $\Delta_0=\|\pmb{\Omega}^0-\pmb{\Omega}^*\|^2$. $\pmb{\Omega}^0$ is the initial model vector, and $\pmb{\Omega}^*$ is the optimal solution of $\mathcal{J}$. $\tilde{\eta}_1=1/\theta_1$, $\theta_1\coloneqq\frac{128L^2\gamma\kappa}{\mu^2}+\frac{96L^2}{\mu}+12(L+\gamma\kappa)$,
 $M_1=E\sigma^2_{F,1} + \sigma^2_{F}/B$, $M_2=E\gamma\kappa\sigma^2_{F,1} + \sigma^2_{F}/B$, and $B$ is the mini-batch size. % $\mathcal{O}(\cdot)$ ignores poly-logarithmic and constant numerical factors.
\begin{IEEEproof}
Due to the limited space, we omit the details of the proof in this paper. We summarize the main idea as follows. By adding a non-linear mapping function to the outer layer of the regularization term, we can prove that the resulting function in (\ref{global_obj}) satisfies new convexity and smoothness conditions. Thus we follow the same line as \cite[Theorem 1]{t2020personalized} to prove ${\cal J}(\pmb{\Omega})$ can converge to the optimum ${\cal J}(\pmb{\Omega}^{*})$ with rate $\mathcal{O}(\frac{1}{T^2 EK})$.
\end{IEEEproof}
\end{theorem}

%\begin{corollary}
%Since model $\pmb{\omega}_{k''}$ is a linear combination of models $\pmb{\omega}_{1}, \cdots, \pmb{\omega}_{k'}$ and the aggregation coefficient $\xi_{k', k''}$ is a constant during model training, $\pmb{\omega}_{k''}$ can inherit the convergence rate of $\pmb{\omega}_{k'}$. Therefore, pSAN can converge on the objective in (\ref{main_objective}) with rate $\mathcal{O}(\frac{1}{T^2 EK})$.
%\end{corollary}

%For illustrative purposes, we compare our rates with those of FL and personalized FL algorithms in i.i.d cases (i.e., $\sigma_{F,1}^2=0$ and $\gamma=1$ ). The strongly-convex rate of {FedWISE} becomes $\frac{\sigma_{F,1}^2}{\mu TN \varepsilon}+\frac{1}{\mu N}$, which matches the lower-bound for the identical case, compared to the latest $\frac{\sigma_F^2}{\mu T N_\tau}+\frac{1}{\mu}$ by SCAFFOLD  and $\frac{\sigma_{F,1}^2}{\mu T N \varepsilon}+\frac{L}{\mu}$ by LSGD-PFL with $L \geq 0$.

\vspace{-0.1cm}
\section{Performance Evaluation}
\subsection{Experiment Setup}
{\bf Dataset:} We use a  public human gesture dataset Widar3 \cite{zheng2019Widar}, which consists of 18 gesture classes performed by 17 human users recorded by 18 Wi-Fi receivers deployed at different locations. Each receiver is equipped with three antennas and can record the CSI measurements at a sampling rate of 1000 Hz. Widar3 covers almost all kinds of sensing scenarios and has been widely used by academia and industry.

{\bf Model:}
%For the convex case, we use an L2-regularization multinomial logistic regression model (MLR) with cross-entropy loss functions. For the non-convex case,
We use a two-layer convolutional neural network (CNN) with two ReLU activation functions, a dropout layer in the middle, and a softmax layer at the end of the network. Besides, we use the SGD optimizer with 0.005 learning rate.
\vspace{-0.1cm}
\subsection{Results}
We first randomly sample 12 receivers from all 18 receivers as {\it Source Receivers} (SRs, including index 1-12), each of which has 18 labeled samples covering 6 types of gestures, while the rest are regarded as the {\it Target Receivers} (TRs, including index 13-18), each of which does not have any labeled gesture samples.
To evaluate the performance of pSAN, we compare the accuracy of pSAN with two traditional model training solutions: local (model) training (Local) and global (model) training (Global) solutions. In local training, each SR uses its own data to train a model in a supervised manner, which may suffer overfitting due to limited local data. In global training, we use FedAvg %is a classical federated learning-based method proposed in \cite{mcmahan2017communication}, which aims at training
to train a single globally shared model based on the datasets of 12 SRs to fit the data distributions of all receivers. In pSAN, all SRs collaboratively train a personalized model for each source one, then we construct the personalized model for each TR by linearly aggregating these uploaded models from SRs based on the semantic similarity between receivers.

\begin{figure}[!ht]
\vspace{-0.3cm}
     \centering
        \includegraphics[width=0.48\textwidth]{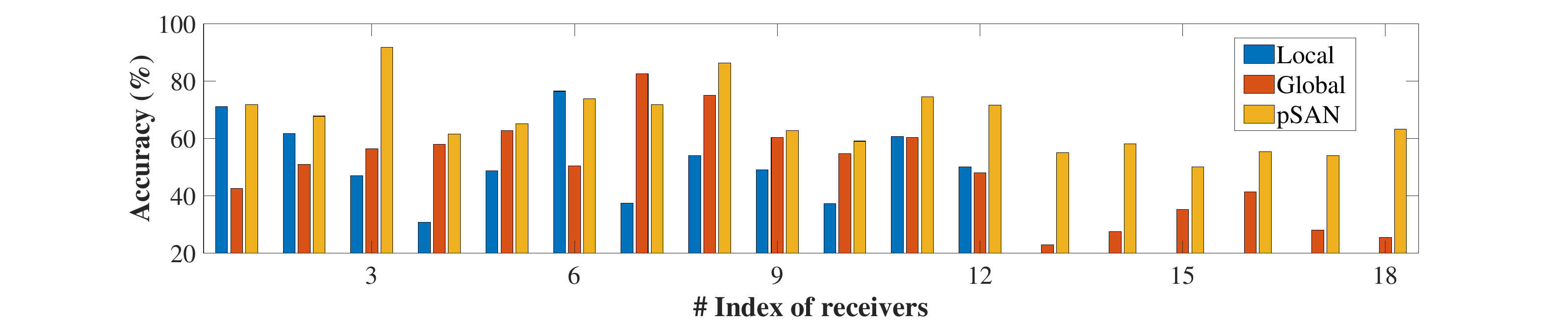}
    \caption{Comparison of local, federated learning-based, and pSAN-based algorithms.}
    \label{Comparison}
\end{figure}
\vspace{-0.3cm}
In Fig. \ref{Comparison}, we observe that pSAN outperforms Local and FedAvg on most SRs (Index 1-12), which means the collaboration-based personalized model training strategy of pSAN can effectively mitigate the issues of label scarcity as well as data heterogeneity at the same time. Besides, we also observe that models of receivers 13-18 constructed by pSAN can offer up to 63.33\% and 37.83\% improvement to Local and Global, respectively.
%This means the proposed semantic-aware personalized model transfer solution is a better choice than the single global model adaptation solution.
From the results, since the FedAvg algorithm relies on a single global model trained on TRs, it is difficult to achieve the desired model accuracy on TRs and SRs due to data heterogeneity in wireless systems. The local training solution also fails to obtain competitive model performance than pSAN due to label scarcity on TRs and SRs. By characterizing the correlation between physical-layer semantic features and the sensing data distributions across different receivers, pSAN customizes the personalized model for each receiver to effectively mitigate the data heterogeneity and label scarcity issues in wireless systems.

\vspace{-0.3cm}
\begin{figure}[!ht]
     \centering
     \begin{subfigure}[b]{0.22\textwidth}
         \centering
           \includegraphics[width=1\textwidth]{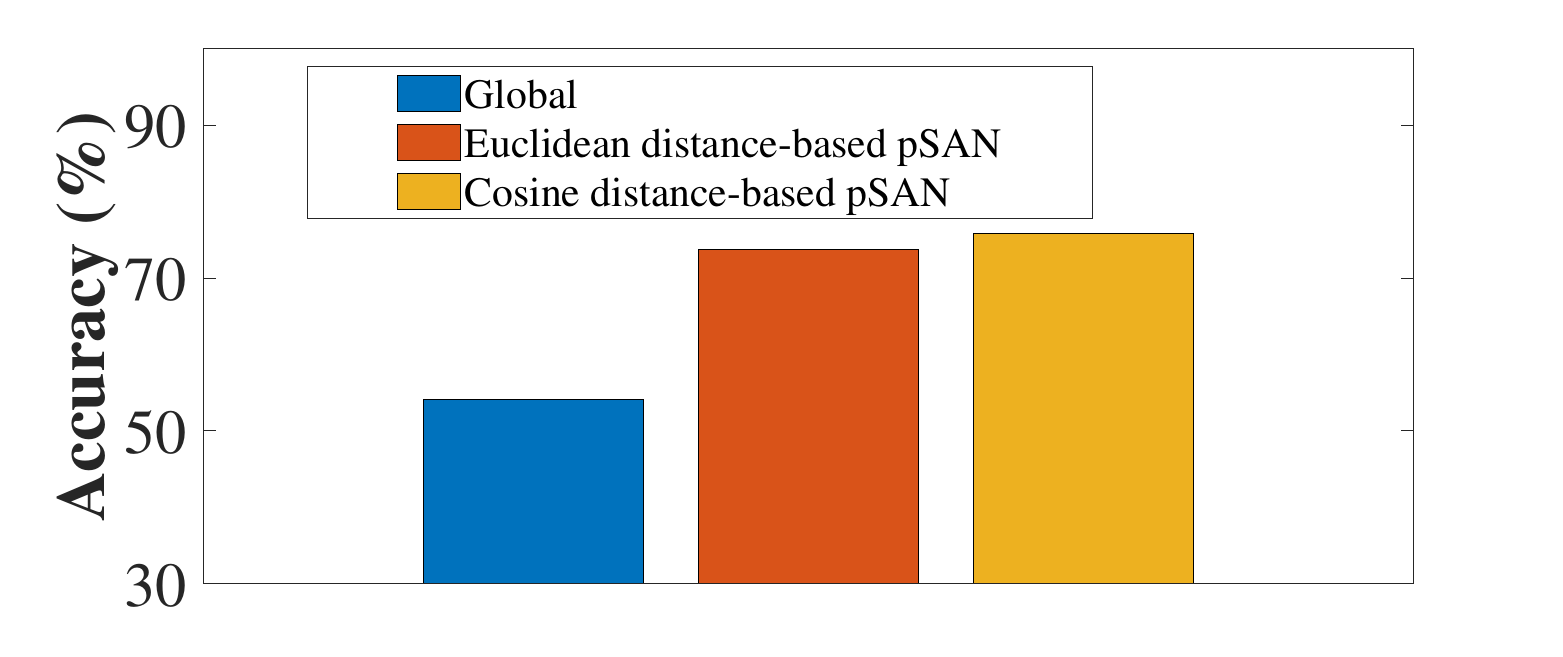}
         \caption{}
         \label{euclidean_cosine}
     \end{subfigure}
     \centering
     \begin{subfigure}[b]{0.22\textwidth}
         \centering
          \includegraphics[width=1\textwidth]{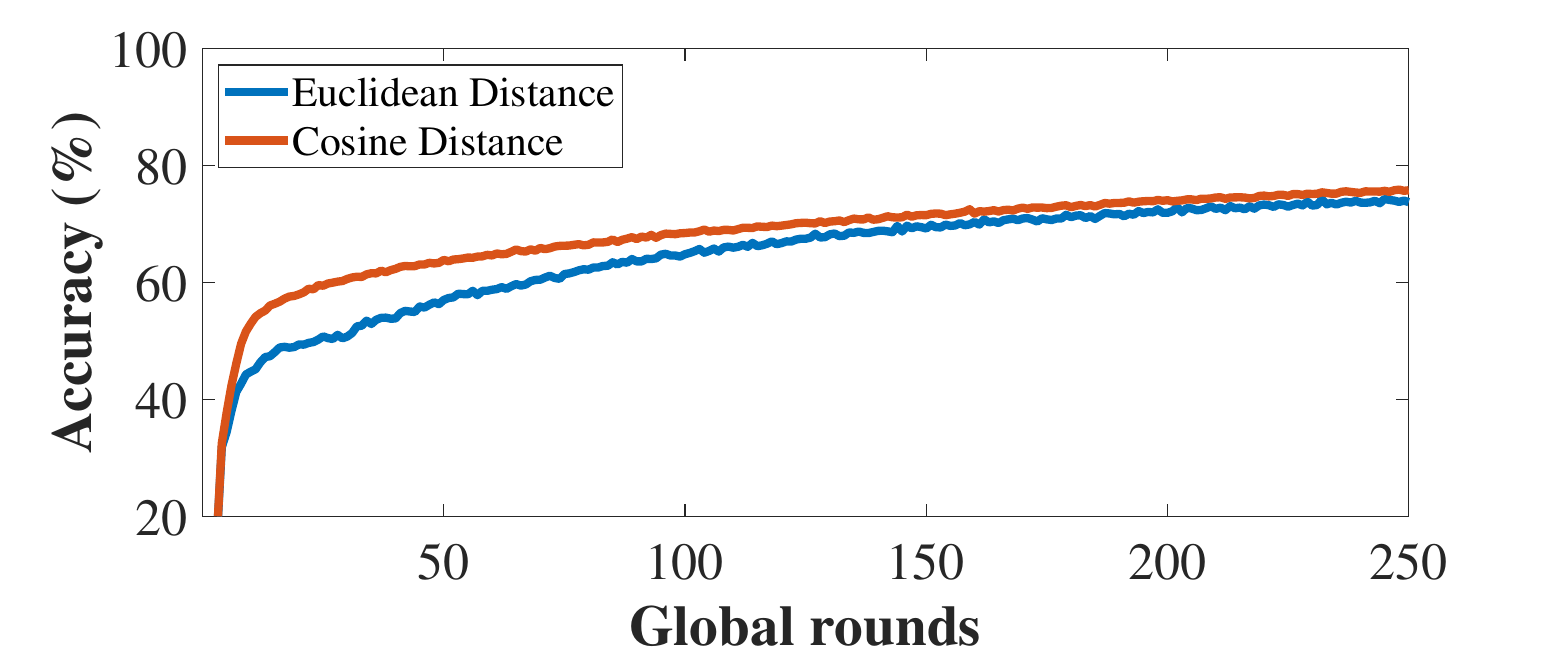}
         \caption{}
         \label{performance_curve}
     \end{subfigure}
    \caption{Comparison of algorithm performance when using different distance metrics.}
    \label{Comparison of algorithm performance when using different distance metrics}
\end{figure}
\vspace{-0.3cm}
%To evaluate the impact of different semantic distance metrics on the ,
In Fig. \ref{Comparison of algorithm performance when using different distance metrics},
we compare the model transfer performance when different distance metrics have been adopted to map semantic distances to model correlations. %  similarity between .
%including the Euclidean and cosine distances to measure semantic similarity and model correlation.
We can observe that, compared to the Euclidean distance, cosine distance achieves a faster convergence rate and higher model accuracy (about 2.02\% improvement with 250 rounds). This is because the Euclidean distance is known to be ineffective in evaluating the difference between two high-dimensional models due to the curse of dimensionality\cite{huang2021personalized}.

\vspace{-0.1cm}
\section{Conclusion}
\vspace{-0.1cm}
This paper proposed a zero-shot wireless sensing solution that allows models constructed in a limited number of locations to be directly transferred to other locations without any labeled data. A novel framework, called pSAN, is developed to characterize the correlation between physical-layer semantic features and the sensing data distributions across different receivers. Then, we propose a pSAN-based zero-shot learning solution in which each receiver can obtain a location-specific gesture recognition model by directly aggregating the already trained models of other receivers. Our results have shown that pSAN offers up to 63.33\% and 37.83\% improvement over local training and global training solutions, respectively.

\vspace{-0.05cm}
\section*{Acknowledgment}
\vspace{-0.1cm}
The work of Y. Xiao was supported in part by the National Natural Science Foundation of China (NSFC) under grant 62071193 and in part by the Key Research \& Development Program of Hubei Province of China under grant 2021EHB015. The work of Y. Li was supported in part by the NSFC under grant 62301516. Y. Xiao, Y. Li, and G. Shi were supported in part by the Major Key Project of Peng Cheng Laboratory under grant PCL2023AS1-2. %, in part , in part by the Key Research and Development Program of Hubei Province under Grant 2021EHB015, and in part by the National Key Research and Development Program of China under Grant 2022YFB2903201.
\vspace{-0.15cm}
\bibliographystyle{ieeetr}
\bibliography{myref,DeepLearningRef}

\end{document}